\documentclass[pre,twocolumn,floatfix,showpacs,preprintnumbers]{revtex4-1}

\usepackage[latin2]{inputenc}
\usepackage{amsmath}
\usepackage{amssymb}
\usepackage{epsfig}
\usepackage{bm}
\usepackage{color}
\usepackage{stackengine}

\begin{document}

\title{Trapping in and escape from branched structures of neuronal dendrites}

\author{Robin Jose}

\author{Ludger Santen}

\author{M. Reza Shaebani}
\altaffiliation{Corresponding author: shaebani@lusi.uni-sb.de}

\affiliation{Department of Theoretical Physics $\&$ Center for Biophysics, Saarland 
University, 66123 Saarbr\"ucken, Germany}

\begin{abstract}
\noindent {\bf ABSTRACT} We present a coarse-grained model for stochastic 
transport of noninteracting chemical signals inside neuronal dendrites 
and show how first-passage properties depend on the key structural 
factors affected by neurodegenerative disorders or aging: the extent 
of the tree, the topological bias induced by segmental decrease of 
dendrite diameter, and the trapping probabilities in biochemical 
cages and growth cones. We derive an exact expression for the 
distribution of first-passage times, which follows a universal 
exponential decay in the long-time limit. The asymptotic mean 
first-passage time exhibits a crossover from power-law to exponential 
scaling upon reducing the topological bias. We calibrate the 
coarse-grained model parameters and obtain the variation range 
of the mean first-passage time when the geometrical characteristics 
of the dendritic structure evolve during the course of aging or 
neurodegenerative disease progression (A few disorders are chosen and 
studied for which clear trends for the pathological changes of dendritic 
structure have been reported in the literature). We prove the validity 
of our analytical approach under realistic fluctuations of structural 
parameters, by comparing to the results of Monte Carlo simulations. 
Moreover, by constructing local structural irregularities, we analyze 
the resulting influence on transport of chemical signals and formation 
of heterogeneous density patterns. Since neural functions rely on 
chemical signal transmission to a large extent, our results open the 
possibility to establish a direct link between the disease progression 
and neural functions.
\end{abstract}
 
\maketitle

\noindent {\bf INTRODUCTION}\\
\noindent The complex behavior of advanced nervous systems mainly 
originates from the elaborate structure of neuronal dendrites 
\cite{Reviews}. The functions of the nervous system substantially 
rely on the diffusion of chemical signals, which is strongly affected 
by the dendrite structure. The branching morphology of dendrites 
allows the neurons to control the transmission time of signals and 
construct a complex network of signaling pathways. While dendritic 
trees share some structural features, e.g.\ branching at acute 
angles or decreasing in their diameter when moving distally from 
soma, their morphology varies widely in different neuronal types 
and regions, reflecting their diverse functions \cite{Waters05}. 
Moreover, the presence of small protrusions along dendrites, 
called spines, adds to the complexity of the system. Spines 
receive excitatory synaptic inputs, temporarily compartmentalize 
them, and undergo dynamic structural changes regulated by neuronal 
activity \cite{SpinesRefs}. Bidirectional communication between 
the spines and the soma (via e.g.\ $\text{Ca}^{2+}$, soluble 
intracellular domains, and subunits of the nuclear import machinery) 
is critical for long-term plasticity, neuronal development, and 
information processing capabilities \cite{BidirectCommunRefs}. 
Additionally, synaptic activation can trigger signaling pathways 
which spread locally in the dendritic channel and influence 
neighboring synapses \cite{ShortRangeTransportRefs}.     

Understanding how signal transmission is governed by the structure
is becoming more important, because pervasive changes of dendritic 
structure have been reported due to aging \cite{Benavides-Piccione13,
Petanjek11,Orner14} or neurodegenerative disorders \cite{DiseaseRefs1,
DiseaseRefs2}, such as Alzheimer's disease \cite{Smith09,AlzheimerRefs1,
AlzheimerRefs2}: (i) the population and spatial extent of branches 
\cite{Orner14,Smith09}, (ii) the thickness, length, and even curvature 
of dendritic channels \cite{Benavides-Piccione13,DiseaseRefs1,Smith09}, 
or (iii) the density, shape, and spatial distribution of spines 
\cite{Benavides-Piccione13,Petanjek11,Orner14,DiseaseRefs1,DiseaseRefs2,
Smith09,AlzheimerRefs1,AlzheimerRefs2} can be affected. To establish 
a link between the structural changes and subsequent alterations of 
neural functions, a deep understanding of the role of structure on 
transport of ions or molecules is still lacking. The attempts 
have been mainly limited to the determination of the impact of 
spine shape on diffusional and first-passage properties of signals 
inside spines \cite{Bloodgood05,Tonnesen14,Takasaki14,Li15,
Santamaria06,Kusters13,Holcman07,Berezhkovskii09}. The role of 
spine density has also been studied by considering comb-like 
structures or (periodically) distributed traps along a channel 
\cite{Dagdug07,Berezhkovskii14,Mendez13,Fedotov08,Bressloff07}. 
However, the precise estimation of escape time from dendritic 
trees to reach soma is a difficult task. The complication arises 
due to complex branching morphology, presence of spines along 
the tree, irregular shape of junctions, and varying cross-section 
radius of dendritic channels. 

\begin{figure}[b]
\centering
\includegraphics[width=0.48\textwidth]{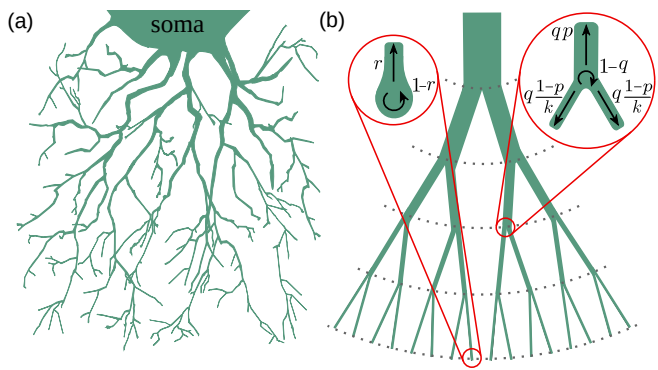}
\caption{(a) Schematic drawing of neuronal dendrites. (b) 
Illustration of the model. An example tree structure with 
$d{=}5$ and $p{=}0.5$ is shown. The arrows indicate 
possible choices at junctions or dead ends, described by 
Eqs.\,(\ref{Eq:MasterEqs}). As a visual guide, the ratio 
between the diameters of parent and child branches is taken 
to be $p{/}(\frac{1{-}p}{k})$ (with $k{=}2$ in dendritic trees).}
\label{Fig1}
\end{figure}

Here we propose a coarse-grained approach to map the stochastic 
transport of ions and molecules inside neuronal dendrites to an 
effective one-dimensional random walk of noninteracting particles 
in a confined geometry. Coarse-grained random walk models have 
been previously employed to successfully describe the influence 
of topological and geometrical characteristics of the structure 
on diffusion in labyrinthine environments (see e.g.\ \cite{Felici04} 
for oxygen absorption in the human lung). Our effective 1D random 
walk model enables us to obtain insightful analytical results 
for mean first-passage times (MFPTs) in complex structures of 
neuronal dendrites. Various types of 1D random walks have been 
previously studied, including biased \cite{Benichou99,Weiss02,
Pottier96,Garcia-Pelayo07} and persistent \cite{Weiss02,Pottier96,
Garcia-Pelayo07,Masoliver89} walks as well as the walks with 
absorption along the path \cite{Benichou99} or at the boundaries 
\cite{Kantor07}. Here, in view of the morphological differences 
between the dendrites of healthy and degenerate brain tissues, 
we concentrate on the major characteristics affected by 
neurodegenerative diseases: the overall extent of dendritic trees, 
the thicknesses of channels, and the structure and density of 
spines. By combining appropriate boundary conditions at the two 
ends of a finite one-dimensional system, partial absorption along 
the path, and biased motion in an effective 1D random walk model, 
we construct a suitable framework to study signal transmission 
in dendrites. We disentangle the contributions of key structural 
features to first-passage properties and verify that the scaling 
behavior of the asymptotic MFPT changes below a threshold value 
of the topological bias induced by hierarchical reduction of 
branch diameter. We evaluate the variation range of the mean 
time required for chemical signals to travel from the synapses 
to the soma in the course of some specific neurodegenerative 
disease progression. Moreover, the applicability of our theoretical 
approach to realistic dendritic structures with spatial heterogeneities 
is addressed and the role of local structural changes on signal 
transmission and formation of heterogeneous density patterns 
is discussed.\\ 

\noindent {\bf METHODS}\\
\noindent By adopting a mesoscopic perspective for transmission 
of ions and molecules inside dendrites, we consider the motion of 
a noninteracting random walker on the nodes of a tree-like regular 
network with a finite depth $d$, parameterizing the extent of 
branches [Fig.\,\ref{Fig1}(b)]. Each node is identified by its 
depth $n$, ranging from $0$ (soma) to $d$ (dead ends). After entering 
the network, the walker randomly jumps to the neighboring nodes until 
it is absorbed in the target, i.e.\ soma. To take into account the 
stochastic trapping events in spines, we assume that the walker either 
moves in the channel or resides inside biochemical cages with probabilities 
$q$ or $1{-}q$, respectively. This way we map the problem to a stochastic 
two-state model. Such models have been widely employed to describe 
altering phases of motion in biological systems \cite{TwoStateModels}. 
Typically, the density of spines (i.e.\ the number of spines per unit 
length along the dendritic channel) quickly saturates after a distance of about 
$50{-}100\,\mu\text{m}$ from soma \cite{Benavides-Piccione13,Ballesteros06,
Rothnie06}. Therefore, we suppose that the residence probability in cages 
is simply depth-independent. The waiting probability at each node is 
an effective measure of the importance of spines in compartmentalizing 
the signals: It increases with increasing the density or head volume 
of spines or decreasing their neck size. To consider the directional 
preference due to e.g.\ hierarchical reduction of branch diameter, a 
topological bias parameter $p$ is introduced for adopting the direction 
of motion at each node. Jumping towards soma or a dead end occurs, 
respectively, with probabilities $p$  or $\frac{1{-}p}{k}$ (with 
$k{=}2$ for the structure of neuronal dendrites). More generally, one 
can adopt a persistent random walk approach \cite{PersistentWalks} to 
include active transport on microtubules or consider passive motion 
in crowded dendritic channels \cite{AntiPRWs}. When arriving at a 
dead end, the walker either returns to the previous junction with 
probability $r$ or explores the connecting channel and the growth 
cone at the tip of the branch with probability $1{-}r$.

\begin{figure}[t]
\centering
\includegraphics[width=0.48\textwidth]{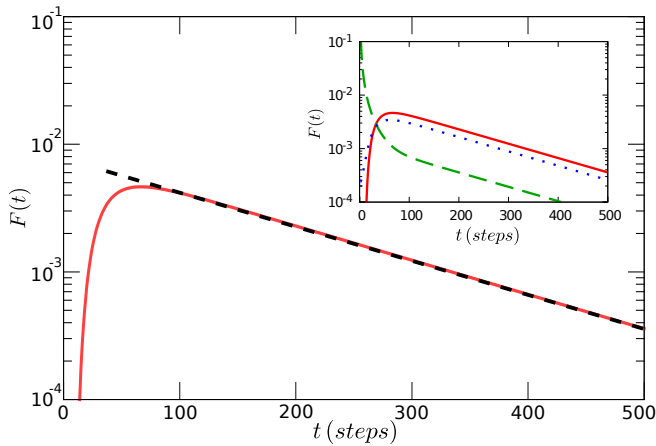}
\caption{First-passage time distribution $F(t)$ for $p{=}q{=}
r{=}\frac12$ and $d{=}10$. The solid line shows the analytical 
result via Eq.\,\ref{Eq:Fz} and the dashed line represents 
the leading exponential term of Eq.\,\ref{Eq:Fz} for $t{\gg}1$. 
Inset: $F(t)$ for the same set of parameter values as in the 
main panel, but for different initial conditions of entering 
the tree. The solid, dashed, and dotted lines correspond to 
the initial conditions $P\!\!_{_n}\!(0){=}\delta_{n,d}$ 
(i.e.\ entering from the dead ends), $P\!\!_{_n}\!(0){=}
\delta_{n,1}$ (entering from the soma), and $P\!\!_{_n}\!
(0){=}(2^{n{-}1})/(2^d{-}1)$ (entering uniformly along the 
tree), respectively.}
\label{Fig2}
\end{figure}

We estimate the mean time required for a particle to escape the 
dendrite structure (characterized by the set of parameters 
$\{d,q,p,r\}$) and reach the soma, by treating the soma as an 
absorbing boundary. However, one can follow the proposed approach 
to investigate the first-passage time for the inverse direction 
(i.e.\ soma-to-spine signaling) as well, by distributing the 
absorbing boundaries along the tree. Let us introduce the 
probability distribution $P\!\!_{_n}\!(t)$ of being at depth 
level $n$ at time step $t$ (In an irregular structure, the 
probability of being at each node can be considered instead). 
The signals initially enter the system via spines, which are 
almost uniformly distributed along dendritic trees. As a result, 
the input rate may even exponentially grow with depth, 
corresponding to the initial condition $P\!\!_{_n}\!(0){=}
\frac{2^{n{-}1}}{2^d{-}1}\,\,(n{\geq}1)$. Here for simplicity 
we consider entering from the dead ends $P\!\!_{_n}\!(0){=}
\delta_{n,d}$, which gives the major contribution to the signal 
input (see the inset of Fig.\,\ref{Fig2}). The analytical 
procedure is however similar for other initial conditions. 
We construct a set of coupled master equations for the 
dynamical evolution of $P\!\!_{_n}\!(t)$ within the framework 
of our stochastic model:
\begin{equation}
\left\{
\begin{array}{ll}
P\!\!_{_0}\!(t) &\!\!\!=\! P\!\!_{_0}\!(t{-}1) {+} 
q \, p \, P\!\!_{_1}\!(t{-}1),\vspace{1mm}\\
P\!\!_{_1}\!(t) &\!\!\!=\! (1{-}q) \, P\!\!_{_1}\!(t{-}1) {+} 
q \, p \, P\!\!_{_2}\!(t{-}1),\\
\;\;\;\;\Shortstack{ . . .} & \\
P\!\!_{_n}\!(t) &\!\!\!=\! q (1{-}p) P\!\!\!_{_{n{-}1}}
\!(t{-}1) {+} (1{-}q) P\!\!_{_n}\!(t{-}1) {+} q p 
P\!\!\!_{_{n{+}1}}\!(t{-}1),\\
\;\;\;\;\Shortstack{ . . .} & \\
P\!\!_{_{d{-}1}}\!(t) &\!\!\!=\! q \, (1{-}p) \, P\!\!_{_{d
{-}2}}\!(t{-}1) {+} 
(1{-}q) \, P\!\!_{_{d{-}1}}\!(t{-}1) {+} r \, P\!\!_{_d}
\!(t{-}1),\vspace{1mm}\\
P\!\!_{_d}\!(t) &\!\!\!=\! q \, (1{-}p) \, P\!\!_{_{d{-}1}}
\!(t{-}1) {+} (1{-}r) \, P\!\!_{_d}\!(t{-}1) {+} \delta(t).
\end{array}
\right.
\label{Eq:MasterEqs}
\end{equation}
The detailed calculations to obtain an expression for the 
escape-time distribution $F(t)$ by solving the above set 
of equations are presented in the {\it Appendix}.\\

\noindent {\bf FIRST-PASSAGE PROPERTIES}\\ 

\noindent The overall shape of the escape-time distribution is 
shown in Fig.\,\ref{Fig2}. Notably, $F(t)$ exhibits an exponential 
tail. We checked that the exponential decay holds independently 
of the choice of the trapping factor $q$, the boundary condition 
$r$ at the deepest branch level, or the chance $p$ of hopping to 
shallower layers. The slope however varies with $q,\,p,\,r$ and 
$d$. Importantly, the inset of Fig.\,\ref{Fig2} shows that while 
the initial conditions of entering the tree may considerably 
influence the overall shape of $F(t)$, the slope of the exponential 
tail remains independent of the way the signals enter the system 
\cite{Tejedor11}. It is technically difficult to extract the tail 
behavior of $F(t)$ from Eq.\,\ref{Eq:Fz} (see {\it Appendix}) in 
general, however, for a given set of parameter values one can deduce 
the exponential asymptotic scaling. The resulting dashed line in 
Fig.\,\ref{Fig2} fully captures the asymptotic slope. As a proof 
of the existence of exponential tail, one can show from Eq.\,\ref{Eq:Fz} 
that the $z$ transform of the first-passage time distribution can 
be written as $F(z){=}\frac{2^{d{+}1}(pqz)^d}{\Phi\!_{_d}(z)}$, 
where $\Phi\!_{_d}(z)$ is a polynomial of maximum degree $d$. By 
evaluating the roots $k$ of the polynomial, it can be verified that 
$F(z){\sim}\frac{1}{\prod\limits_{k{=}1}^{d^*}(1{-}\alpha
\!_{_k} z)^{\beta\!_{_k}}}$, where $\alpha\!_{_k}$ is a function 
of the structural parameters and $d^*\!,\beta\!_{_k}{\leq}d$. 
Then, after partial fraction decomposition of $F(z)$ and inverse 
$z$ transform, $F(t)$ can be represented as a sum of $\alpha
\!_{_k}^{\,t}$ terms, thus, can be approximated by the leading 
exponential term $\alpha\!_{_{k,\text{max}}}^{\,t}$ in the limit 
$t{\rightarrow}\infty$.

The mean-first-passage time $\langle\,\!t\rangle$ of chemical 
signals to reach the soma, which is our main quantity of interest, 
can be evaluated from $F(t)$ as explained in details in the {\it 
Appendix}. The analytical Eq.\,\ref{Eq:MET} in the {\it Appendix} 
represents the mean-first-passage time in terms of the coarse-grained 
model parameters. Although the expression is continuous, it is 
indeterminate at $p{=}\frac12$. By taking the limit we get 
$\langle\,\!t\rangle{=}\frac{qd-rd+rd^2}{qr}$ for the specific 
choice $p{=}\frac12$.

\begin{figure}[t]
\centering
\includegraphics[width=0.48\textwidth]{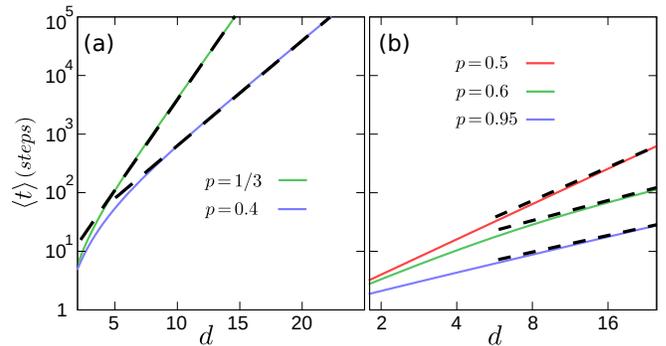}
\caption{Mean escape time vs the depth of the tree for (a) 
$p{<}\frac12$ (log-lin scales) and (b) $p{\geq}\frac12$ 
(log-log scales) at $q{=}r{=}1$. The analytical results 
of Eq.\,\ref{Eq:MET} are shown with solid lines, and the 
dashed lines represent the asymptotic exponential or power-law 
scaling of $\langle\,\!t\rangle$ via Eq.\,\ref{Eq:METasymp}.}
\label{Fig3}
\end{figure}

\noindent {\it Crossover in asymptotic scaling behavior--} 
To clarify how $\langle\,\!t\rangle$ varies with the model 
parameters, we exclude the indeterminate point  $p{=}\frac12$ 
to simplify the MFPT expression. For $p{\neq}\frac12$, 
Eq.\,\ref{Eq:MET} of the {\it Appendix} reduces to the 
sum of a linear and an exponential function of $d$,
\begin{equation}
\left< t \right> = \displaystyle\frac{d}{q\,(2p-1)} + 
\frac{p\, r - p \, q\,(2p-1)}{q\,r\,(2p-1)^2}\Big((
\frac{1}{p}-1)^d {-}1\Big).
\label{Eq:MET2}
\end{equation}
Hence, $\langle\,\!t\rangle$ in the limit $d{\gg}1$ scales 
exponentially (linearly) for $0{<}p{<}\frac12$ ($\frac12{<}
p{<}1$), as the exponential term on the right-hand side of 
Eq.\,\ref{Eq:MET2} dominates (vanishes). It can be also seen 
that $\langle\,\!t\rangle$ for the specific choice $p{=}
\frac12$ scales as a power-law $d^\gamma$ with $\gamma{=}2$. 
Thus, the crossover of the asymptotic mean escape time from 
a power-law to an exponential scaling can be summarized as
\begin{equation}
\langle t \rangle \sim
\begin{cases} 
\displaystyle \frac{1}{q\,(2p{-}1)}\,d,  
& \;\;\;\;\;\;\;\; 1/2{<}p{<}1, \vspace{1mm}\\
\displaystyle q^{-1} \, d^2,  & \;\;\;\;\;\;\;\; 
p{=}1/2, \vspace{1mm}\\
\displaystyle\frac{p\, r {-} p \, q\,(2p{-}1)}{q\,r\,
(2p{-}1)^2} \, e^{d \ln\left(\frac{1}{p}-1\right)},  
& \;\;\;\;\;\;\;\; 0{<}p{<}1/2.
\end{cases}
\label{Eq:METasymp}
\end{equation}
The first (last) case indeed grows logarithmically (linearly) 
with the number of nodes in regularly branched trees 
\cite{DendrimerRef}. Figure\,\ref{Fig3} indicates that 
the asymptotic slopes are properly captured by the analytical 
prediction of Eqs.\,\ref{Eq:METasymp}. The change in the 
scaling behavior of $\langle\,\!t\rangle$ from linear to 
exponential at the threshold value $p_c{=}\frac12$ in a 
1D random walk can be understood because the effective 
direction of flow (with respect to the target) is inverted. 
This also induces a transition from recurrent to transient 
random walks in infinite trees \cite{BiasedBetheLattice}. 
At $p{=}\frac12$, the balance between the two directions 
of diffusive transport holds and it is expected that the 
bias parameter $p$ in a healthy neuron is around this 
threshold value. \\

\noindent {\bf RESULTS AND DISCUSSION}\\
\noindent {\bf Coarse-grained model calibration}\\ 
In the following, we compare our analytical results to those 
obtained from ordinary diffusion at microscopic scales in 
dendritic spines and other relevant geometries such as 
thickening tubes, to verify the applicability of our 
coarse-grained approach and to calibrate the model 
parameters. Note that the diffusion problem with a 
constant diffusion coefficient across the structure 
is basically a linear differential equation. If the trend 
of the first-passage time versus one of the model parameters 
or as a function of a related geometrical characteristic of 
actual dendrite structures match, then our model parameter 
can be calibrated into the dendritic structure through 
a fit to micro-scale computations for pure diffusion.

The structure of neuronal dendrites primarily depends on 
the nervous system and varies in different neuronal regions 
and cell types. However, as a reference for comparison, here we 
have chosen typical cerebellar Purkinje cells of guinea pigs 
which extend nearly $200\,\mu\text{m}$ from the soma and have 
${\sim}\,450$ dendritic terminals \cite{Rapp94}. Thus, there 
are nearly $10$ generations of junctions in such a structure 
(corresponding to $d{=}10$ in our coarse-grained view) and they 
branch out every $20\,\mu\text{m}$ on average.

\begin{figure*}[t]
\centering
\includegraphics[width=0.95\textwidth]{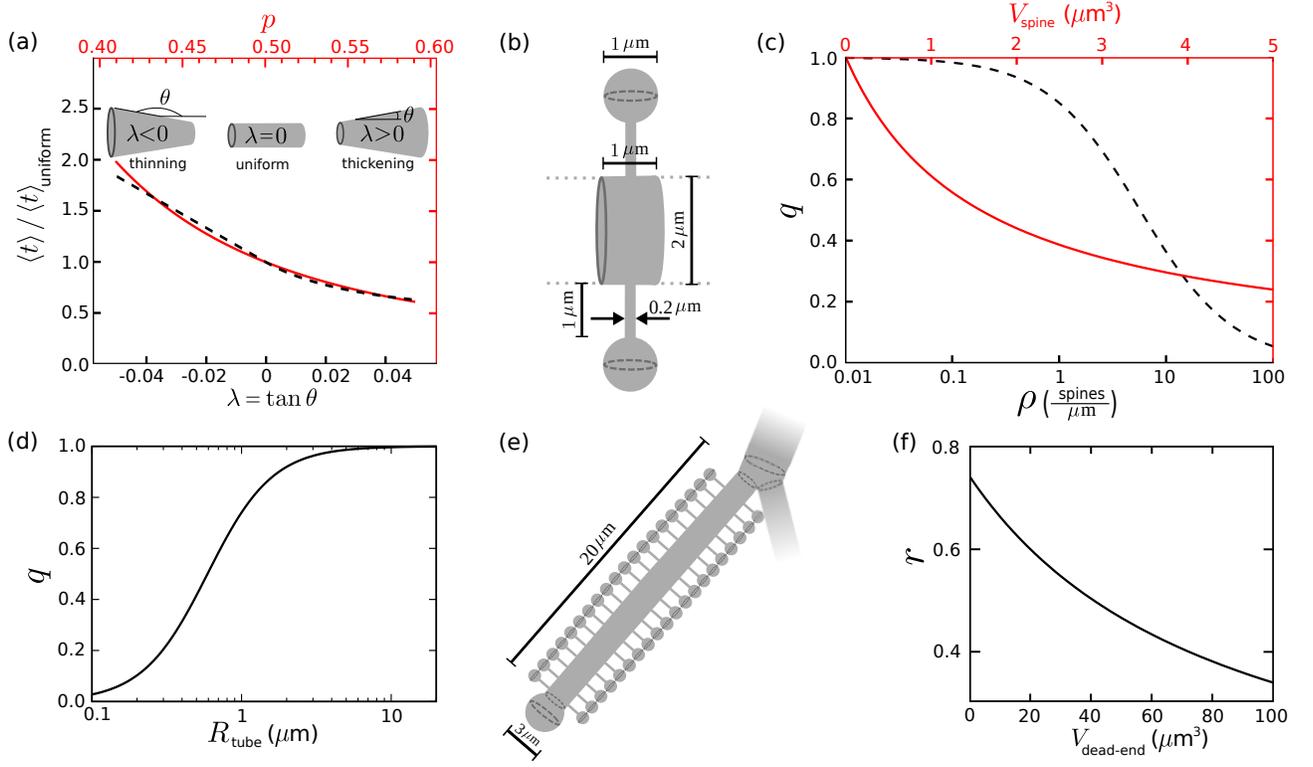}
\caption{(a) The mean-first-passage time between the two ends 
of a tube, scaled by the result of a uniform tube, versus 
$\lambda{=}\tan\theta$ or the bias parameter $p$. The results 
of Eqs.\,\ref{Eq:MET} and \ref{Eq:MFPT-tube} are shown with 
solid and dashed lines, respectively. Inset: Schematic drawing 
of tubes of varying cross section. (b) Typical size scales taken 
as the reference values in healthy dendrites. (c) The parameter 
$q$ versus the spine density $\rho$ (dashed line) or the spine 
volume $V_\text{spine}{=}V_\text{head}{+}V_\text{neck}$ (solid 
line). (d) $q$ vs the radius of the dendritic tube. (e) The 
regularized geometry of the segment of the dendritic tube 
connecting the dead end to the last branch point. (f) The 
coarse-grained parameter $r$ in terms of the volume of the 
dead end. The parameter values (unless varied) are taken to 
be $\rho{=}2\,\frac{\text{spines}}{\mu\text{m}}$, 
$R_{_\text{tube}}{=}1\,\mu\text{m}$, $V_\text{spine}{\simeq}
0.55\,\mu\text{m}^3$, and $L_{_\text{tube}}{=}20\,\mu\text{m}$ 
in panels (c), (d) and (f).}
\label{Fig4}
\end{figure*}

{\it --\,The bias parameter $p$:} The problem of diffusion in a 
tube of varying cross section has been thoroughly studied both 
theoretically and numerically in the literature \cite{RefsTube}. 
Particularly, Brownian dynamics simulations at microscopic scales 
were employed in \cite{Berezhkovskii07} to explore the range of 
validity of an effective one-dimensional description of diffusion 
in uniformly thickening or thinning tubes. Denoting the opening 
angle of the tube with $\theta$ [see the inset of Fig.\,\ref{Fig4}(a)], 
they approximated the mean-first-passage time $\langle\,\!t\rangle$ 
between the two ends of a tube of length $L_{_\text{tube}}$ and 
initial radius $R_{_\text{tube}}$, scaled by $\langle\,\!t\rangle$ 
of a tube of uniform cross section, as
\begin{equation}
\langle t \rangle{/}\langle t \rangle\!_{_\text{uniform}} \simeq
\begin{cases} 
\displaystyle \frac{\sqrt{1{+}\lambda^2}}{3}\,(3{+}2\lambda\,
\tilde{L}),  
& \;\;\;\;\;\;\;\; \lambda{<}0, \vspace{3mm}\\
\displaystyle\frac{\sqrt{1{+}\lambda^2}}{3}\,\frac{3{+}\lambda
\,\tilde{L}}{1{+}\lambda\,\tilde{L}},  
& \;\;\;\;\;\;\;\; 0{<}\lambda,
\end{cases}
\label{Eq:MFPT-tube}
\end{equation}
where $\lambda{=}\tan\theta$ and $\tilde{L}{=}L_{_\text{tube}}{/}
R_{_\text{tube}}$. Their analytical and simulation results match 
for opening angles $\theta{<}10^\circ$. Even such small thickening 
rates are still larger than what is typically observed in neuronal 
dendrites. For example, the thickness of the dendritic channel 
varies from nearly $0.5$ around the dead ends to less than $8\,
\mu\text{m}$ close to soma in cerebellar Purkinje cells of guinea 
pigs which has a typical extent of $200\,\mu\text{m}$, i.e.\ an 
opening angle of less than $2^\circ$ \cite{Rapp94}. Therefore, 
within the validity range of their analytical expressions, we 
compare $\langle\,\!t\rangle$ obtained from our coarse-grained 
approach Eq.\,\ref{Eq:MET} to their results in Fig.\,\ref{Fig4}(a). 
We set $q{=}r{=}1$ to avoid trapping since Eq.\,\ref{Eq:MFPT-tube} 
is valid for smooth tubes with reflecting walls. We also consider 
our reference dendritic structure [see Figs.\,4(b),(e)] for ease 
of comparison. The scaled MFPTs obtained via Eqs.\,\ref{Eq:MET} and 
\ref{Eq:MFPT-tube} fit very well using a simple linear map between 
$\lambda$ and $p$ as $\lambda{\sim}\,0.5p{-}0.25$. We checked that, 
within biologically relevant parameter ranges and weakly thickening 
regime $\theta{<}5^\circ$, one can obtain similar satisfactory 
agreement between the mean first-passage times by treating the 
coefficients of the linear transformation as fit parameters. In 
the following, we choose $p{=}0.55$ as the reference value for 
our coarse-grained bias parameter in a typical healthy dendrite 
(corresponding to $\theta{\simeq}1.4^\circ$). Note that the 
geometry of the junctions may affect the first passage results 
in general, however, we expect that it causes minor variations 
since the cross-section area at the branch point is conserved.

\begin{figure*}[t]
\centering
\includegraphics[width=0.95\textwidth]{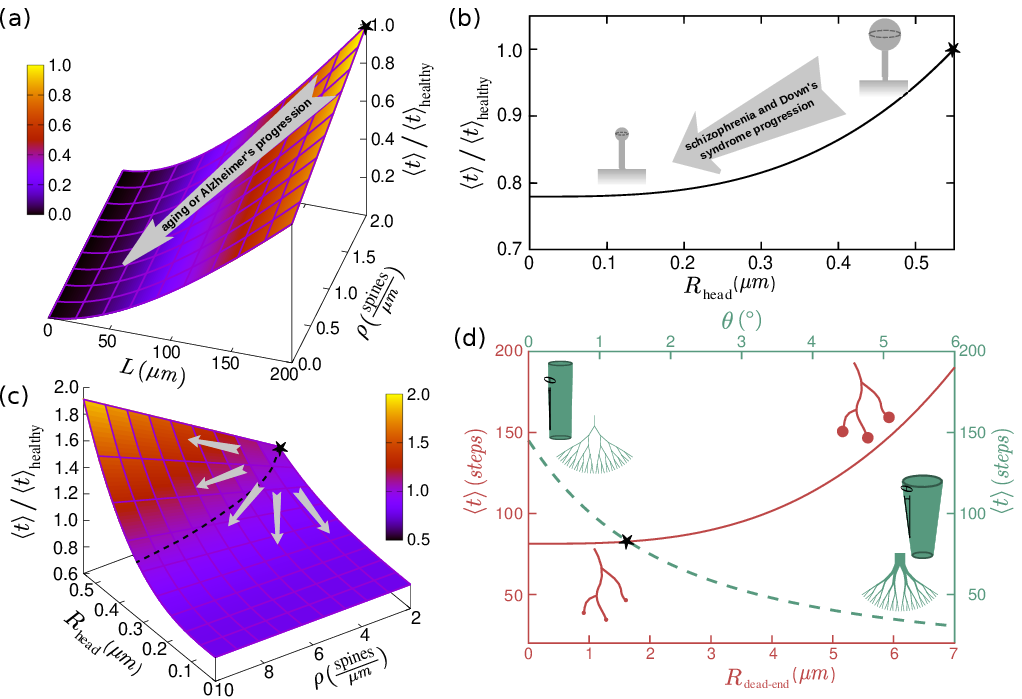}
\caption{Influence of the pathologies of spines and dendrites 
on chemical signal transmission. The mean first-passage time 
$\langle\,\!t\rangle$, scaled by the MFPT of the reference 
healthy structure $\langle\,\!t\rangle_{\text{healthy}}$, 
versus the structural characteristics affected in the course 
of (a) aging and Alzheimer's disease (spine density $\rho$ and 
extent of the dendritic tree $L$), (b) schizophrenia and Down's 
syndrome (spine head radius $R_{_\text{head}}$), and (c) 
fragile X syndrome ($\rho$ and $R_{_\text{head}}$). The 
stars mark the corresponding point for the reference 
healthy structure. In panel (c), the arrows represent the 
possible directions of fragile X progression, and the dashed 
contour line marks the path along which $\frac{\langle
\,\!t\rangle}{\langle\,\!t\rangle_{\text{healthy}}}{=}1$. 
(d) Variation of the MFPT with changing the opening angle 
$\theta$ of the dendritic tube (dashed line) or the radius 
$R_{_\text{dead-end}}$ of the spherical growth cones (solid 
line).}
\label{Fig5}
\end{figure*}

\noindent {\it --\,The trapping parameter $q$:} The coarse-grained 
parameter $q$ in our model indeed represents the fraction of time 
spent in the dendritic channel in the steady state, which is set 
by the probabilities $\kappa_w$ and $\kappa_m$ of switching from 
motion in the channel to waiting in the spines and vice versa. 
$\kappa_w$ is proportional to the density of spines and the 
mean entrance area of the spine neck and inversely proportional 
to the cross-section area of the dendritic channel. Thus, one 
obtains $\kappa_w{\propto}\rho\,R_{_\text{neck}}^2{/}
R_{_\text{tube}}^2$, where $R_{_\text{neck}}$, $\rho$ and 
$R_{_\text{tube}}$ denote the neck radius, spine density and 
radius of the dendritic channel, respectively. $\kappa_m$ is 
inversely proportional to the mean escape time from spines 
$\langle\,\!t\rangle_{_\text{spine}}$, which obeys \cite{Berezhkovskii09}
\begin{equation}
\langle\,\!t\rangle_{_\text{spine}} = \frac{L^2_\text{neck}}{2D} 
{+} \frac{L^2_\text{neck} V_\text{head}}{D\,V_\text{neck}} {+} 
\frac{V_\text{head}}{4 \, D \, R_\text{neck}},
\label{Eq:METspines}
\end{equation}
with $D$ being the diffusion coefficient and $L_\text{neck}$, 
$V_\text{neck}$ and $V_\text{head}$ denoting, respectively, 
the neck length and volume and the head volume of the spines. 
The diffusion coefficient depends on the size of the diffusing 
object. For example, the typical value of $D$ in dendritic 
spines for $\text{Ca}^{2+}$ ions and green fluorescent 
protein (GFP) variants (that are much smaller in size) were 
reported to be ${\sim}100$ and $20\,\mu\text{m}^2{/}
\text{s}$, respectively \cite{Yasuda11}. Inside the dendrite 
channel, $D\sim37\,\mu\text{m}^2{/}\text{s}$ was obtained 
for a specific photoactivatable GFP (paGFP) \cite{Bloodgood05}. 
Similar results were reported for diffusion in other cell 
types. For comparison, $D$ was found to be ${\sim}23.5$ 
and $25.2\,\mu\text{m}^2{/}\text{s}$ for the motion of 
enhanced GFP (eGFP) inside the nucleus and in the cytoplasm 
of HeLa cells, respectively \cite{Chen02}. For a typical 
thin spine \cite{Harris92} with $R_{_\text{neck}}{=} 100 
\, \text{nm}$, $L_\text{neck}{=}1\,\mu\text{m}$, and a 
head diameter of $1\,\mu\text{m}$ (thus with $V_\text{neck}{
\simeq}0.03\,\mu\text{m}^3$ and $V_\text{head}{\simeq}0.52
\,\mu\text{m}^3$) as shown in Fig.\,\ref{Fig4}(b), one 
gets $\langle\,\!t\rangle_{_\text{spine}}{\simeq} 0.19$, 
$0.48$, and $0.77\,\sec$ for the escape time of 
$\text{Ca}^{2+}$, paGFP, and eGFP from spines 
(using $D\!_{_\text{Ca}}{=}{100}$, $D\!_{_\text{paGFP}}
{=}{40}$, and $D\!_{_\text{eGFP}}{=}{25}\,\mu\text{m}^2
{/}\text{s}$). Moreover, the mean travel time of signals 
from synapses to soma in smooth dendritic channels of 
length $x$ can be estimated from Eq.\,\ref{Eq:METspines} 
as $t{\simeq}\frac{x^2}{2D}$. By choosing $x{=}20\,\mu\text{m}$ 
as an example, one obtains $2.0$, $5.0$, and $8.0\,\sec$ 
for the travel time of $\text{Ca}^{2+}$, paGFP, and 
eGFP, respectively.

The transitions between the two states of motility are 
non-Markovian in general, however, one can estimate the 
asymptotic value of $q$ in the limit $t{\rightarrow}\infty$ 
as a function of the volumes of the dendritic tube and 
spines as \cite{Dagdug07}
\begin{equation}
q=\displaystyle\frac{\kappa_m}{\kappa_m{+}\kappa_w}\simeq
\displaystyle\frac{V_\text{tube}}{V_\text{tube}{+}V_
\text{spines}}=\displaystyle\frac{1}{1{+}\displaystyle
\frac{\rho}{\pi\,R_{_\text{tube}}^2}(V\!\!_{_\text{head}}
{+}V\!\!_{_\text{neck}})}.
\label{Eq:q-asymptotic}
\end{equation}
Figure\,\ref{Fig4}(c) shows how the $q$ parameter varies with 
the spine density and volume. While increasing $\rho$ or 
$V_\text{spine}$ enhances the trapping probability and thus 
reduces $q$, increasing the volume of the dendritic tube 
leads to longer excursion times in the tube and increases 
$q$, as shown in Fig.\,\ref{Fig4}(d). By choosing $\rho{=}
2\,\frac{\text{spines}}{\mu\text{m}}$, $R_{_\text{tube}}{=}
1\,\mu\text{m}$, and $V_\text{head}{+}V_\text{neck}{\simeq}
0.55\,\mu\text{m}^3$ \cite{Harris92}, we obtain the healthy 
reference value $q{\simeq}0.74$ for further comparisons.

\noindent {\it --\,The boundary-condition parameter $r$:} 
Finally, we calibrate the parameter $r$ via a similar 
procedure as explained for $q$. The coarse-grained 
parameter $r$ effectively represents the probability 
of motion inside the segment of the dendritic tube which 
connects the last branch point to the dead end [see the 
schematic Fig.\,\ref{Fig4}(e)]. By ignoring the minor 
corrections due to the negligible thickening along such 
a short tube segment, the asymptotic value of $r$ can 
be approximated as
\begin{align}
r&{\simeq}\displaystyle\frac{V_\text{tube}}{V_\text{tube}
{+}V_\text{spines}{+}V_\text{dead-end}} \\ \nonumber
&=\displaystyle\frac{1}{1{+}\displaystyle
\frac{\rho}{\pi\,R_{_\text{tube}}^2}(V_\text{head}{+}
V_\text{neck}){+}\frac{V\!\!_{_\text{dead-end}}}{\pi\,
R_{_\text{tube}}^2\,L_{_\text{tube}}}}.
\label{Eq:r-asymptotic}
\end{align}
Let us consider a spherical dead end with a typical diameter 
of $3\,\mu\text{m}$ and assume that the tree branches out 
every $20\,\mu\text{m}$ on average. Then, using the rest of 
the reference parameter values used for the determination of 
the $q$ parameter, we get $r{\simeq}0.63$. The variation 
of $r$ as a function of the volume of the dead end is shown 
in Fig.\,\ref{Fig4}(f). Even in the absence of the dead end 
(i.e.\ $V\!\!_{_\text{dead-end}}{=}0$), the signals may 
be still trapped in the spines distributed between the dead 
end and the last junction, leading to $r{\neq}1$.\\

\noindent {\bf Influence of pathological changes on transmission 
of chemical signals}\\ 
After adopting the set of model parameter values $p{=}0.55$, 
$q{=}0.7$, $r{=}0.6$ and $d{=}10$ as the reference for healthy 
structures of dendrites, next we investigate how far the 
mean-first-passage time varies when the geometrical characteristics 
of the dendritic structure evolve during the course of aging or 
neurodegenerative disease progression. Here we choose aging and 
a few examples of neurodegenerative disorders (such as Alzheimer's 
disease, schizophrenia, and fragile X and Down's syndromes), 
for which, clear trends for the pathological changes of dendritic 
structure have been reported in the literature \cite{Fiala02}. 
In the course of aging or Alzheimer's progression, both the 
density of spines and the extent of the dendritic tree reduce 
\cite{AlzheimerRefs1,Giannakopoulos09,Benavides-Piccione13} (The 
spine density of the apical dendrites of pyramidal neurons in the 
cingulate cortex of humans may decrease to less than $60\%$ with 
aging \cite{Benavides-Piccione13}). These changes are equivalent 
to the increase of $q$ and reduction of $d$ in our coarse-grained 
perspective. It is also known that the schizophrenia and Down's 
syndrome progression leads to the reduction of the spine size 
\cite{Roberts96,Marin-Padilla72}, corresponding to the enhancement 
of our $q$ parameter. The pathology of fragile X makes the 
prediction of MFPT variations complicated. In the course of 
fragile X progression, while the spine density increases 
(enhancement of $q$), their shapes become more elongated 
and the spine head volume reduces (reduction of $q$) 
\cite{He13,Irwin01,Wisniewski91}. Therefore, we expect that 
the variation of $q$ (and thus of the MFPT) is less pronounced 
in fragile X compared to the other examples. The competition 
between the variations of spine density and shape determines 
whether $q$ effectively decreases or increases in the course 
of fragile X progression.

In Fig.\,\ref{Fig5}, we show the trends of the MFPTs upon 
changing the dendritic structure due to aging or diseases, 
as explained above. The combined effects of the reduction 
of tree extent and spine density due to aging or Alzheimer's 
disease can dramatically decrease the MFPT of chemical signals 
from the synapses to the soma [Fig.\,\ref{Fig5}(a)]. To 
calculate the MFPT, we used Eq.\,\ref{Eq:MET} with $q$ 
inserted from Eq.\,\ref{Eq:q-asymptotic} and $d{=}L\,(\mu
\text{m})/20$. The reduction of both spine density and tree 
extent to half of their healthy reference values decreases 
the relative MFPT to $\frac{\langle\,\!t\rangle}{\langle\,
\!t\rangle_{\text{healthy}}}{\simeq}0.3$. Thus, the system 
gradually loses the ability to compartmentalize ions and 
molecules and maintain chemical concentrations to a wide 
extent. The shrinkage of the spine size in schizophrenia 
and Down's syndrome leads to a similar trend for the variation 
of MFPT, however, the effect is less pronounced. In the extreme 
case of zero head volume, the MFPT reduces to nearly $80\%$ of 
its reference value (see panel (b) of Fig.\,\ref{Fig5}). As a 
result of the competition between the increase of spine density 
(up to $\rho{\approx}10\,\frac{\text{spines}}{\mu\text{m}}$) 
and reduction of spine head volume (down to $R_{_\text{head}}
{=}0$) in fragile X syndrome, the relative MFPT, $\frac{\langle
\,\!t\rangle}{\langle\,\!t\rangle_{\text{healthy}}}$, may vary 
within the range of $[0.6, 1.9]$. If the reduction rate of 
spine head volume equals the growth rate of spine density, 
the two effects compensate each other and the MFPT remains 
unchanged, as shown by the contour line in Fig.\,\ref{Fig5}(c). 
In other neurodegenerative disorders, the pathology of spine 
and dendrite structure is more complicated. For example, 
distortion of spine shape in most mental retardations 
\cite{DiseaseRefs2} makes the prediction of the MFPT trend 
difficult. Another point is that there is currently a lack 
of quantitative studies to clarify the impact of diseases 
or aging on the thickening of dendritic tubes (corresponding 
to the variation of our $p$ parameter) or on the 
morphological changes of growth cones (variation of $r$). In 
Fig.\,\ref{Fig5}(d), we calculate the MFPTs within reasonable 
variation ranges of the opening angle of the dendritic tube 
or the radius of spherical growth cones. Here we use 
Eq.\,\ref{Eq:MET} with $q$ inserted from Eq.\,9 and $p$ 
from the linear relation $\tan(\theta){\simeq}\,0.5p{-}0.25$. 
One obtains up to 3-fold increase or reduction in $\langle
\,\!t\rangle$ compared to the healthy reference $\langle\,\!t
\rangle_{\text{healthy}}$.

The mean travel time of chemical signals in dendrites
reflects the ability to preserve local concentrations 
or induce concentration gradients of ions and molecules, 
thus, it is tightly connected to neural functions. 
Therefore, the quantitative evaluation of the first-passage 
times in different diseases is a step forward towards 
linking the disease progression to neural functions 
and draw physiological conclusions.\\

\begin{figure*}[t]
\centering
\includegraphics[width=0.97\textwidth]{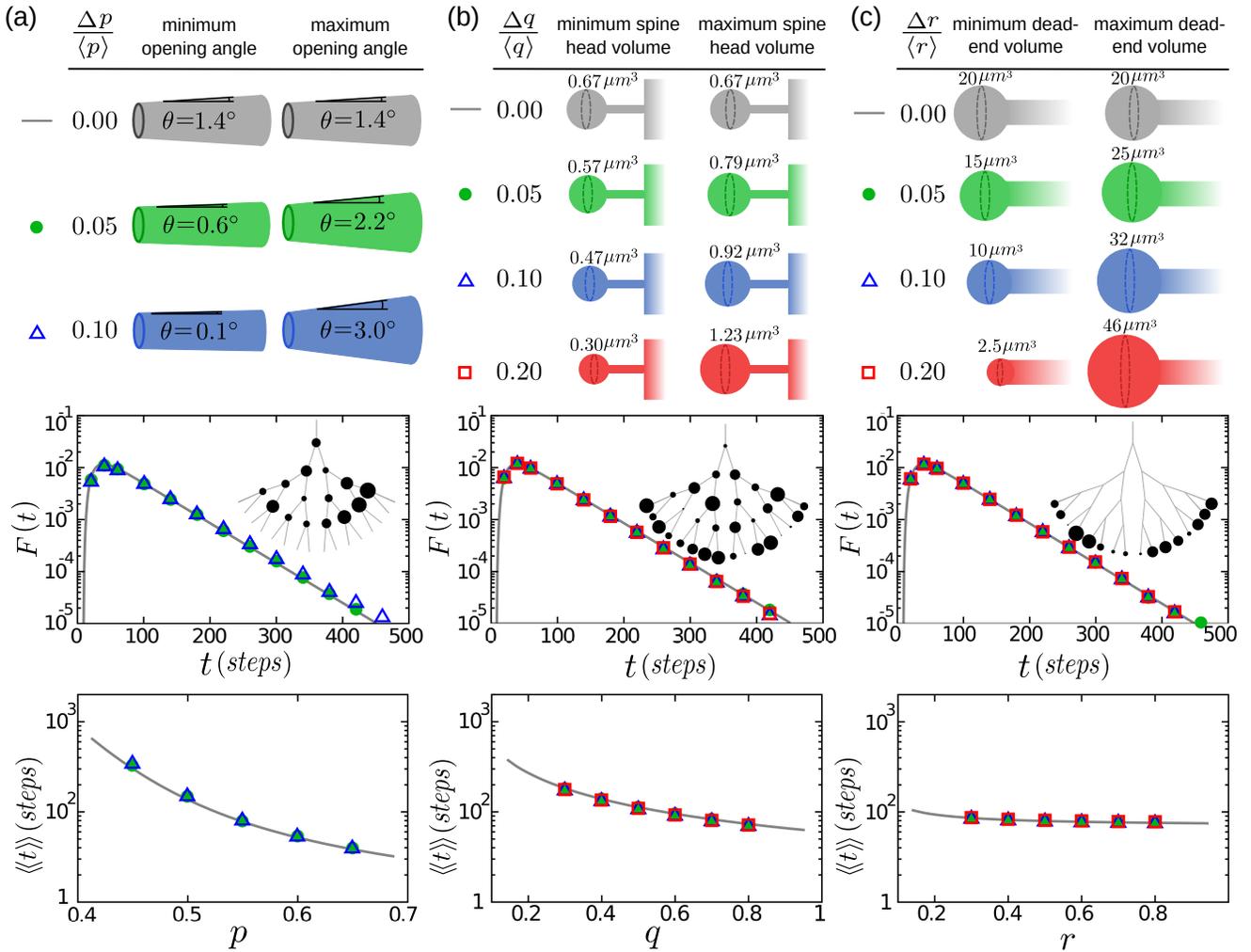}
\caption{Comparison between the analytical predictions for 
constant parameter values and simulation results for dynamically 
varying (a) $p$, (b) $q$, or (c) $r$ parameter across the 
dendritic structure. The reference parameter values are taken 
to be $p{=}0.55$, $q{=}0.7$, $r{=}0.6$ and $d{=}10$. Upper 
panels: Schematic representation of the variations in the 
realistic dendritic geometry when the coarse-grained model 
parameters vary $5$, $10$ or $20\%$ around their mean reference 
values. Middle panels: Log-lin plots of the escape-time 
distribution. The analytical curve via Eq.\,\ref{Eq:Fz} 
(solid line) is compared to the simulation results (symbols). 
The insets are schematic diagrams of typical trees with 
$d{=}5$ and $10\%$ fluctuations in the corresponding model 
parameter. The radii of the circles are proportional to 
the relative deviations from the minimum values. Lower panels: 
Mean escape time versus the model parameters. The solid line 
represents the analytical prediction of Eq.\,\ref{Eq:MET} and 
the symbols correspond to the simulation results. $\langle\!
\langle{\cdot}{\cdot}{\cdot}\rangle\!\rangle$ denotes averaging 
over both the ensemble of realizations for a given disorder 
and the ensemble of possibilities for the stochastic particle 
dynamics.}
\label{Fig6}
\end{figure*}

\noindent {\bf Structural irregularities}\\ 
In our analytical formalism, we consider constant coarse-grained 
parameters along the entire tree. This corresponds to the 
assumption of a spatially homogeneous structure, i.e.\ if the 
dendritic tree is regularly branched, the channels thicken with 
the same rate throughout the tree, the spine density and size 
are spatially uniform, and all the growth cones are of the same 
size. From a coarse-grained perspective, such a regular structure 
can be described by a few major parameters, which allows for the 
calculation of the MFPTs. Taking into account that our 
coarse-grained parameters indeed represent the key structural 
features which undergo pathological changes in the course of 
neurodegenerative disease progression, the model enables us 
to connect the disease progression to signal transmission, 
as discussed in the previous section. However, realistic 
dendritic structures are spatially heterogeneous. For example, 
the density of spines may vary even up to $40\%$ around the 
global mean value in dendritic trees \cite{Benavides-Piccione13,
Ballesteros06}. The spines also undergo dynamic structural 
changes regulated by neuronal activity \cite{SpinesRefs}. 

In view of the realistic structural fluctuations, the 
basic question is whether the analytical predictions via our 
coarse-grained approach remain valid when the structural 
parameters of a given dendritic tree are allowed to spatially 
vary around their global mean values. In the following, 
we compare the analytical result for the reference set of 
parameter values with the simulation results where the 
structural parameters spatially fluctuate around the 
reference values. For comparison, the fluctuation range 
$\frac{\Delta\,q}{\langle q\rangle}{\approx}0.2$ is 
comparable to the realistic variations in spine head size 
and density in pyramidal neurons in the singulate cortex 
of humans \cite{Benavides-Piccione13}. Similar fluctuation 
ranges are considered for $d$, $r$ and $p$ parameters, in 
the absence of quantitative studies to explore the variation 
ranges of the extent of dendritic trees, the size of growth 
cones, and the thickening rate of dendritic channels.

In each of the Monte Carlo simulations, we vary only one of 
the coarse-grained parameters while the rest of them are fixed 
at their mean values $\langle d \rangle{=}10$, $\langle p 
\rangle{=}0.55$, $\langle q \rangle{=}0.7$, or $\langle r 
\rangle{=}0.6$. Let us first consider the parameters $p$, 
$q$, and $r$. A new value is assigned to the variable 
parameter at each random walk step, which is randomly 
taken from a uniform distribution in the interval 
$[p{-}\Delta\,p, p{+}\Delta\,p]$, $[q{-}\Delta\,q, 
q{+}\Delta\,q]$, or $[r{-}\Delta\,r, r{+}\Delta\,r]$ for 
parameter $p$, $q$, or $r$, respectively. The upper panels 
of Fig.\,\ref{Fig6} represent the variation ranges of the 
geometrical characteristics of dendritic structures as the 
width of the uniform distributions for coarse-grained 
parameters vary from $0$ up to $20\%$ around the reference 
(healthy) values. In the upper panel of Fig.\,\ref{Fig6}(b) 
we present the extreme values of the spine head volume as 
a function of $\frac{\Delta\,q}{\langle q\rangle}$. However, 
one can alternatively fix the head volume (e.g.\ at 
$V\!\!_{_\text{head}}{=}1\,\mu\text{m}^3$) and consider 
the changes in the spine density and get $[1.3, 1.3]$, 
$[1.1, 1.6]$, $[0.9, 1.8]$, and $[0.6, 2.5]$ intervals 
for the number of spines per micron at $\frac{\Delta\,q}{
\langle q\rangle}{=}0,\,0.05,\,0.1,\,\text{and}\,0.2$, 
respectively. The middle panels show that the resulting 
escape-time distributions $F(t)$ invisibly deviate from 
the analytical prediction (solid line) for $q$ and $r$ 
parameters, while the tail of $F(t)$ starts deviating from 
the theory line when $p$ varies up to $10\%$ around $\langle 
p \rangle{=}0.55$. However, such tail deviations have an 
insignificant impact on the mean-first-passage time 
$\langle\,\!t\rangle$, as shown in the lower panel of 
Fig.\,\ref{Fig6}(a) at $p{=}0.55$. We also repeated the 
simulations for other sets of reference parameters to 
check whether $\langle\,\!t\rangle$ deviates from the 
analytical prediction. According to the results shown 
in the lower panels of Fig.\,\ref{Fig6}, we conclude 
that our analytical results are robust against realistic 
fluctuations of the structural characteristics across the 
dendritic trees (even up to $20\%$ around the mean), 
over a wide range around the reference set of 
coarse-grained parameter values.

\begin{figure}[t]
\centering
\includegraphics[width=0.47\textwidth]{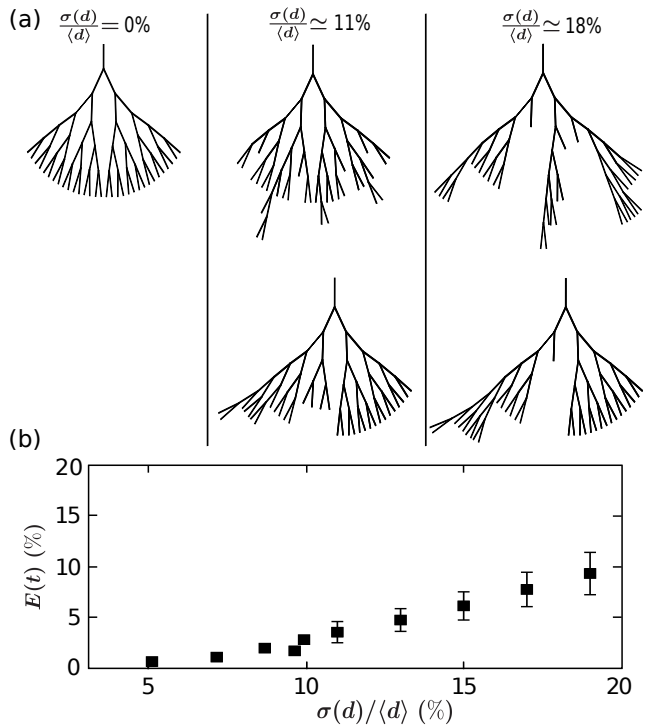}
\caption{(a) Schematic drawings of typical heterogeneous 
trees with $32$ dead ends and a global variation of the 
order of $\frac{\sigma(d)}{\langle d \rangle}{\simeq}11
\,\text{or}\,18\%$ in their extent. The lower trees are 
examples of highly asymmetric structures, with a 
regularly-branched right wing and an irregular left wing. 
(b) The deviation of the analytical MFTP $\langle\,\!t
\rangle\!_{_\text{theory}}$ (calculated for the average 
depth $\langle d \rangle$) from the simulation result 
$\langle\!\langle\,\!t\rangle\!\rangle$, characterized 
by $E(t){=}\frac{|\langle\!\langle\,\!t\rangle\!\rangle
{-}\langle\,\!t\rangle\!_{_\text{theory}}|}{\langle\!
\langle\,\!t\rangle\!\rangle}$, in terms of the fluctuation 
range of the dead-end depths. $\langle\!\langle{\cdot}{\cdot
}{\cdot}\rangle\!\rangle$ denotes averaging over both the 
ensemble of configurations for a given disorder and the ensemble 
of possibilities for the stochastic particle dynamics. 
The parameter values are taken to be $p{=}0.55$ and 
$q{=}r{=}1$. The data points without error bars represent 
single realizations while those with error bars are averages 
of $E(t)$ over bins of size $2$ along the $x$-axis. Error 
bars represent the standard deviation of $E(t)$.}
\label{Fig7}
\end{figure}

Next, we investigate the variations in the extent of the tree 
around the mean value $\langle d \rangle$. To this aim, in 
Monte Carlo simulations we construct stochastic tree structures 
by randomly allowing the nodes to have their child nodes in a 
hierarchical manner starting from the root node. The procedure 
continues until the tree consists of a given number of dead ends. 
A few examples of the resulting structures with $32$ dead ends 
are shown in Fig.\,\ref{Fig7}(a). We characterize the depth of 
the irregular tree by the average of its dead-ends depths $\langle 
d \rangle$, and its variation by $\frac{\sigma(d)}{\langle d 
\rangle}$, with $\sigma(d)$ being the standard deviation. As 
shown in Fig.\,\ref{Fig7}(a), the ensemble of structures 
corresponding to a given $\frac{\sigma(d)}{\langle d \rangle}$ 
contains globally heterogeneous configurations as well as highly 
asymmetric ones. In Fig.\,7(b), we show how the deviation from 
our analytical prediction grows with increasing the fluctuation 
range of the dead-end depths. It can be seen that the error of the 
analytical expression remains below $10\%$ even in considerably 
heterogeneous structures with $\frac{\sigma(d)}{\langle d \rangle}{
\approx}20\%$. For lower variations in the extent of the tree 
($\frac{\sigma(d)}{\langle d \rangle}{<}10\%$), the error is 
less than $5\%$. Thus, our analytical approach is applicable to 
dendritic structures with moderate heterogeneity in their branching 
pattern.

\begin{figure*}[t]
\centering
\includegraphics[width=0.97\textwidth]{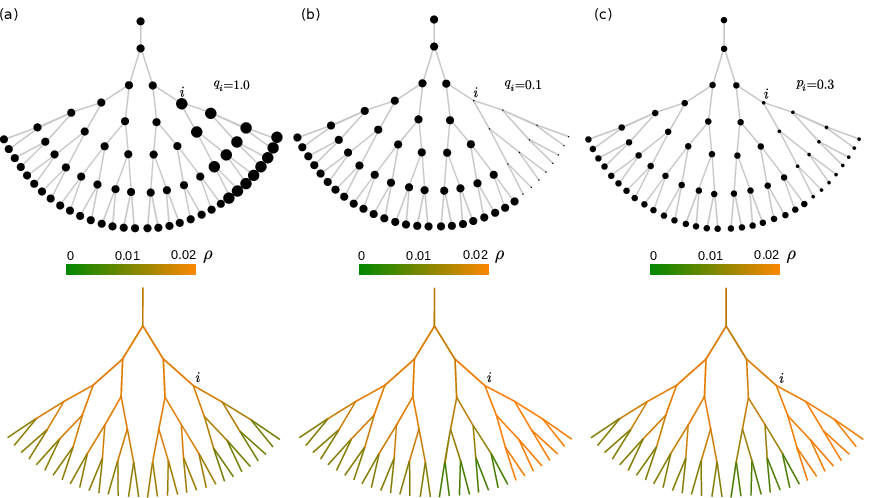}
\caption{Upper panels: Schematic diagrams of example trees 
with local structural irregularities in the sub-branch 
starting from junction $i$. Common parameters (unless 
locally varied): $d{=}6,\;p{=}0.55,\;q{=}0.7,\;\text{and}
\,r{=}0.6$. The modified coarse-grained parameter in the 
sub-branch is denoted by $q_{_i}$ or $p_{_i}$. The radii 
of the circles are proportional to the local $q$ (a,b) 
or $p$ (c) values. Lower panels: Heat maps of the stationary 
density of particles in each branch in simulations performed 
for the corresponding asymmetric structures.}
\label{Fig8}
\end{figure*}

So far, we have investigated the influence of global 
dynamic irregularities of structure on the first-passage 
times. However, static local irregularities may also 
exist in real dendrites, induced by pathological changes. 
For example, various dendritic abnormalities associated 
with fibrillar amyloid deposits in transgenic mouse model 
of Alzheimer's disease and in human brain were reported in 
\cite{Grutzendler07}. Extensive spine density loss, shaft 
atrophy (i.e.\ decline in the radius of the dendritic tube), 
and formation of varicosity (which consists of an enlarged 
tortuous and crumpled part of dendritic channel) were observed 
in the vicinity of amyloid deposits. From our coarse-grained 
perspective, the local varicosity formation, shaft atrophy 
and spine-density reduction correspond, respectively, to 
the decrease of $q$ and $p$ and increase of $q$ in a specific 
region of the tree such as a sub-branch. To elucidate the 
impact of static local irregularities on signal transmission, 
our effective 1D analytical description based on the depth 
levels does not help. Therefore, we construct the entire tree 
structure in Monte Carlo simulations again (similar to the 
procedure to construct asymmetric configurations in Figure\,\ref{Fig7}, 
but this time for $q$ or $p$ parameters). Figure\,\ref{Fig8} 
shows a few samples of coarse-grained dendritic trees with 
an affected sub-branch. We change $q$ or $p$ parameter in 
the affected sub-branch, while keeping the rest of parameters 
the same as the entire tree. By imposing a constant entrance 
rate (one particle from one of the randomly chosen dead 
ends at each time step), we eventually obtain the spatial
distribution of non-interacting particles in the steady 
state. The results shown in Fig.\,\ref{Fig8} reveal that 
local structural irregularities influence the transport 
of particles and lead to the formation of heterogeneous 
density patterns in the system. Reduction of $q$ or $p$ 
in the affected sub-branch to $q_{_i}{=}0.1$ or $p_{_i}{=}0.3$ 
imposes a local trap and leads to a local population 
which is, respectively, $43\%$ or $28\%$ higher than 
the homogeneous case. On the other hand, increasing the 
local $q$ to $q_{_i}{=}1$ reduces the population in the 
sub-branch to $88\%$ compared to the regular tree. Such 
uneven distributions of signaling ions and molecules may 
have dramatic consequences on neural activities such as 
neuronal firing and the ability to maintain chemical 
concentrations and gradients.\\

\noindent {\bf CONCLUSION AND OUTLOOK}\\
\noindent In summary, an analytical framework has been 
developed to obtain first-passage times of chemical signals 
in neuronal dendrites in terms of the structural factors 
which undergo pathological changes in the course of 
neurodegenerative disease progression. By quantitatively 
connecting the dendritic structure to signal transmission, 
our results open the possibility to establish a direct link 
between the disease progression and neural functions, which 
allows to draw important physiological conclusions. 

To consider structural inhomogeneities and dynamical variations 
of real dendrites, the master equations 1 can be generalized 
by introducing uncorrelated probability distributions for the 
key structural parameters $p$, $q$, and $r$ and calculating 
the first-passage properties in terms of their first two 
moments. The fluctuation of depth $d$ can be also taken 
into account by distributing the dead-end conditions among 
the master equations which belong to a given range of the 
deepest levels of the tree. Moreover, the MFPT of passive 
particles in crowded dendritic channels or active ones along 
microtubules \cite{Newby09} can be taken into account in our 
master equations by introducing a (anti-)persistent random 
walker. The interparticle interactions at high density 
regimes affect the transport through the narrow necks 
of spines, which influences the waiting time distribution 
in spines and the first-passage properties. The investigation 
of these aspects calls for additional research efforts. 
The proposed approach provides an analytical route into 
a variety of search and transport phenomena on complex 
networks (e.g.\ weighted time-varying trees), branched 
macromolecules and polymers, various energy landscapes, 
and more generally biased random walks with absorbing 
boundaries \cite{OutlookRefs}. Our calculations can be 
adapted to real labyrinthine environments by introducing 
node-degree distribution and closed paths.\\

\noindent {\bf Appendix: first-passage time calculations}\\
To derive an expression for the first-passage time distribution, 
we start from the master Eqs.\,\ref{Eq:MasterEqs} and obtain 
a set of equations for $P\!\!_{_n}\!(z)$ at different depth 
levels by defining the $z$ transform $P\!\!_{_n}\!(z){=}
\sum\limits_{t{=}0}^{\infty} P\!\!_{_n}\!(t)\,z^t$ (with 
$|z|{<}1$). For example, the transformation of the last 
equation in the set of master Eqs.\,\ref{Eq:MasterEqs} 
reads $P\!\!_{_d}\!(z) \!= q \, (1{-}p) \, z \, P\!\!_{_{d
{-}1}}\!(z) {+} (1{-}r) \, z \, P\!\!_{_d}\!(z) {+} 1$, where 
the constant term results from the $z$ transform of the delta 
function. The challenge is that the number of equations $d{+}1$ 
is arbitrary. However, after some algebra, we solve this set 
of equations to obtain $P\!\!_{_n}\!(z)$, from which the $z$ 
transform of the first-passage time distribution to reach 
the soma can be evaluated as $F(z){=}\sum\limits_{t{=}0}^\infty 
F(t) z^t{=}q\,p\,z\,P\!\!_{_1}\!(z)$ \cite{Redner01}. Let 
us define $\lambda_{\pm}{=}\frac{1}{qpz}\big[1{+}(q-1)z{\pm}
A(z,q,p)\big]$, where $A(z,q,p)=\displaystyle\Big[1{+}2(q{-}1)
z{+}\big[1{-}2q{+}(1{-}2p)^2\,q^2\big]z^2\Big]^{1{/}2}$. We 
derive the following exact expression for the $z$ transform 
of the escape-time distribution, 
\begin{widetext}
\begin{equation}
F(z) =  \displaystyle \frac{2^{d+1} \, r \, A(z,q,p)}{\big(
\lambda_+^d - \lambda_-^d\big) \left[ r\Big(1-\big(q-1\big)z\Big) 
+ p \, q\big({-}2+2 \, z -r \, z\big) \right] + \big(\lambda_+^d 
+ \lambda_-^d\big) r \, A(z,q,p)}.
\label{Eq:Fz}
\end{equation}
\end{widetext}
Next, by inverse $z$ transforming of $F(z)$, one gets an explicit 
lengthy expression for $F(t)$ in terms of the number of time steps 
$t$. We confirmed the correctness of our calculations by comparing 
the analytical predictions via Eq.\,\ref{Eq:Fz} to the results of 
extensive Monte Carlo simulations obtained from $10^6$ realizations 
of the same stochastic process. The mean-first-passage time 
$\langle\,\!t\rangle$ can be evaluated as $\langle\,\!t\rangle
{=} \displaystyle\sum_{t{=}0}^\infty t \, F(t) {=}z 
\frac{d}{d z}F(z)\Bigr|_{z{\rightarrow}1}$. By expanding 
Eq.\,\ref{Eq:Fz} around $z{=}1$ up to first order terms, 
$F(z){\sim}F(z)\Bigr|_{z{\rightarrow}1}{+}(z{-}1)\frac{d}{d z}F(z)
\Bigr|_{z{\rightarrow}1}{+}\mathcal{O}\Big((z{-}1)^2\Big)$, 
and defining $\gamma_{\pm}{=}\frac{1}{p}\big(1{\pm}|2p{-}1|\big)$ 
we arrive at the following expression for the mean escape time,
\begin{widetext}
\begin{equation}
\langle t \rangle = 2^{d{+}1} \frac{\big(\gamma_-^d {-} 
\gamma_+^d\big) \Big[\big(2 \, p \, q{+}d \, r\big)\big(
1{-}2 \, p\big)+2 \, p \, r\Big] + \big(\gamma_-^d {+} 
\gamma_+^d\big) d \, r \, |2p{-}1|}{q \, r \Big[ \big(
\gamma_-^d {-} \gamma_+^d\big) \big(1{-}2p\big) + \big(
\gamma_-^d {+} \gamma_+^d\big) \, |2p{-}1|\Big]^2}\,
\Theta(1{-}2p),
\label{Eq:MET}
\end{equation}
\end{widetext}
where $\Theta(x){=}\{\begin{smallmatrix}+1 \;\;\;0{\leq}
x\\-1\;\;\;x{<}0\end{smallmatrix}$. In the limit $d{\rightarrow}
\infty$, $\langle\,\!t\rangle$ diverges as expected for 
infinite Cayley trees \cite{Redner01} and Bethe lattices 
\cite{BetheLatticeRefs,BiasedBetheLattice}. \\

\noindent {\bf AUTHOR CONTRIBUTIONS}\\
\noindent Correspondence and request for materials should be 
addressed to M.R.S.\ (email:\ shaebani@lusi.uni-sb.de). L.S. 
and M.R.S. designed the research. R.J. and M.R.S. performed 
the research. All authors contributed to the analysis and 
interpretation of the results. M.R.S. wrote the manuscript.\\

\noindent {\bf ACKNOWLEDGEMENTS}\\
\noindent This work was funded by the Deutsche Forschungsgemeinschaft 
(DFG) through Collaborative Research Center SFB 1027 (Projects 
A7 and A8).


\begin{thebibliography}{10}
\bibitem{AntiPRWs} The effective persistency of the motion can 
be characterized by the turning-angle distribution $P(\theta)$ of the particle. 
Introducing $\epsilon{=}\!\int\!\text{d}\theta\,\cos(\theta)\,P(\theta)$, one 
obtains a negative $\epsilon$ for motion in a crowded environment where 
the particle is frequently reflected from obstacles and experiences sharp 
turns (i.e.\ a higher chance of motion to the backward directions), while gets 
a positive $\epsilon$ for active motion along microtubules, where the 
walker continues along the previous direction of motion unless when it 
switches to another filament with opposite polarity. In (anti-)persistent
random walks, the stochastic equations of motion are generalized by introducing 
the parameter $\epsilon$ to take the previous direction of motion into account.
\end{thebibliography}
\end{document}